\documentclass[aip, reprint]{revtex4-2} 

\usepackage[T1]{fontenc}
\usepackage{times}
\usepackage{graphicx}
\usepackage{booktabs}
\usepackage{amsfonts, amsmath,amssymb, bm, graphicx}
\usepackage{siunitx}
\usepackage[english]{babel}
\usepackage{physics}
\usepackage{changes}
\usepackage[colorlinks=true,allcolors=blue]{hyperref}
\usepackage{footmisc}
\usepackage{upgreek}
\usepackage{soul}

\newcommand{\figref}[2]{\hyperref[#1]{\ref{#1}(#2)}}
\newcommand{\figrefsub}[3]{\hyperref[#1]{\ref{#1}(#2)#3}}

\makeatletter
\let\ORIbbl@fixname\bbl@fixname
\def\bbl@fixname#1{%
  \@ifundefined{languagealias@\expandafter\string#1}
    {\ORIbbl@fixname#1}
    {\edef\languagename{\@nameuse{languagealias@#1}}}%
}
\newcommand{\definelanguagealias}[2]{%
  \@namedef{languagealias@#1}{#2}%
}
\makeatother

\definelanguagealias{en}{english}

\newcommand{\HZDR}{Helmholtz-Zentrum Dresden--Rossendorf, Institut f\"ur Ionenstrahlphysik und Materialforschung, D-01328 Dresden, Germany}
\newcommand{\TUD}{Fakult\"at Physik, Technische Universit\"at Dresden, D-01062 Dresden, Germany}
\newcommand{\CNN}{Centre de Nanosciences et de Nanotechnologies, CNRS, Universit\'e Paris-Saclay, 91120 Palaiseau, France}
\newcommand{\GF}{GlobalFoundries Dresden Module One LLC \& Co. KG (GF), 01109 Dresden, Germany}

\begin{document}

\title{Electrical magnon spectroscopy with foundry-fabricated nanoscale magnetic tunnel junctions}

\author{Christopher Heins}
\affiliation{\HZDR}
\affiliation{\TUD}

\author{Zeling Xiong}
\affiliation{\HZDR}
\affiliation{\TUD}

\author{Attila K\'akay}
\affiliation{\HZDR}

\author{Joo-Von Kim}
\affiliation{\CNN}

\author{Thibaut Devolder}
\affiliation{\CNN}

\author{Aleksandra Titova}
\affiliation{\GF}

\author{Johannes M\"uller}
\affiliation{\GF}

\author{Ren\'e H\"ubner}
\affiliation{\HZDR}

\author{Andreas Worbs}
\affiliation{\HZDR}

\author{Ryszard Narkowicz}
\affiliation{\HZDR}

\author{J\"urgen Fassbender}
\affiliation{\HZDR}
\affiliation{\TUD}

\author{Katrin Schultheiss}\email{k.schultheiss@hzdr.de}
\affiliation{\HZDR}

\author{Helmut Schultheiss}\email{h.schultheiss@hzdr.de}
\affiliation{\HZDR}

\date{\today}


\begin{abstract}

Magnonics uses spin waves and their quanta, magnons, to carry and process information and offers new opportunities for computing and signal processing. Its integration with conventional electronics, however, requires scalable electrical interfaces for magnon readout. Here, we demonstrate electrical magnon spectroscopy using a 70-nm magnetic tunnel junction fabricated on a commercial 300-mm semiconductor production line. Dynamic dipolar fields generated by magnons in a magnetic vortex drive the free layer of the tunnel junction and are transduced into electrical signals through tunnelling magnetoresistance. By exploiting three-magnon splitting, we spectrally separate the magnon response from direct microwave excitation and resolve nonlinear magnon modes with linewidths down to 188 kHz at 3.59 GHz. The electrical measurements reveal spectral structure that remains unresolved in parallel Brillouin light scattering measurements. Our results establish foundry-fabricated magnetic tunnel junctions as nanoscale electrical interfaces for magnonic systems and provide a route towards their integration with semiconductor electronics.

\end{abstract}

\maketitle


\begin{figure*}
    \includegraphics{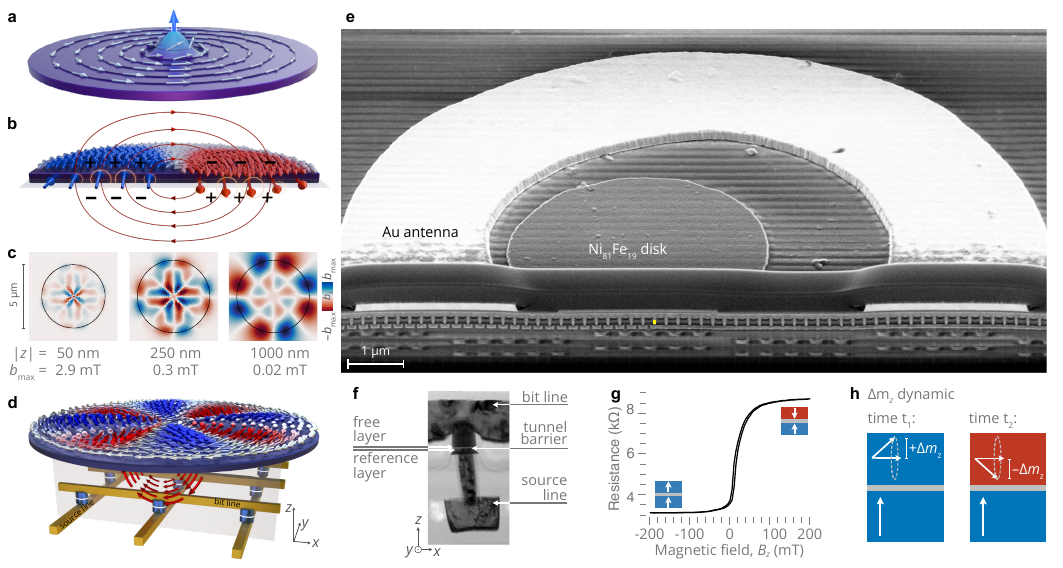}
     \caption{ \textbf{Foundry-fabricated MTJ platform for electrical magnon spectroscopy.} 
     \textbf{a}, Magnon cavity consisting of a 50-\SI{}{\nano\meter}-thick, 5-\SI{}{\micro\meter}-diameter Ni$_{81}$Fe$_{19}$ disk in the magnetic vortex state. The magnetic moments align predominantly in the sample plane and form concentric rings around the vortex core, where the magnetisation points out of the plane. \textbf{b}, Schematic illustration of the dynamic dipolar fields generated by magnons excited in a vortex-state disk for the ($0,1$) mode. The precessional motion of the magnetic moments creates surface charges that oscillate in time at the magnon frequency and in space on the scale of the magnon wavelength. \textbf{c}, Micromagnetic simulation of the $z$-component of the dipolar fields generated by the ($0,4$) mode at different distances $|z|$ from the disk surface. The \SI{250}{\nano\meter} distance approximately corresponds to the experimental separation between the magnetic disk and the MTJ free layer. \textbf{d}, Hybrid structure combining the magnon cavity with a customised MRAM array. \textbf{e}, Scanning electron microscopy (SEM) image of a focused-ion-beam cross-section of the hybrid structure, recorded at an angle of 52$^\circ$. The lower part shows a row of customised MRAM cells connected by bit and source lines, while the upper surface shows the magnetic disk surrounded by an $\Omega$-shaped microwave antenna for magnon excitation. \textbf{f}, Transmission electron microscopy (TEM) image of a customised MTJ fabricated by GlobalFoundries, Dresden. \textbf{g}, Magnetoresistance of a customised MTJ as a function of out-of-plane magnetic field, with the effective perpendicular magnetic anisotropy modified by free-layer interface engineering. \textbf{h}, Schematic of the dynamic response of the free layer. With its equilibrium magnetisation in the sample plane, the magnon-induced dipolar field drives a precessional motion that produces an oscillating tunnelling magnetoresistance.
    }
    \label{fig:platform}
\end{figure*}

Magnons, the quanta of the collective spin-wave excitations in magnetic systems, offer a means to process and transmit information using the electron spin rather than its charge\cite{vvkruglyakMagnonics2010,chumakMagnonSpintronics2015,chumakAdvancesMagneticsRoadmap2022, flebus2024MagnonicsRoadmap2024}. Their wavelike properties can be harnessed for signal-processing operations such as Fourier transforms\cite{csabaSpinwaveBasedRealization2014,csabaPerspectivesUsingSpin2017} and majority-gate logic\cite{fischerExperimentalPrototypeSpinwave2017, talmelliReconfigurableSubmicrometerSpinwave2020}, while their nonlinear dynamics enable computational tasks such as pattern recognition\cite{pappNanoscaleNeuralNetwork2021, korberPatternRecognitionReciprocal2023} and time-series forecasting~\cite{xiongPredictingFutureMagnons2026}. With characteristic frequencies in the GHz regime and wavelengths extending into the deep-submicrometre range, magnons in ferromagnets operate on spatial and temporal scales compatible with conventional semiconductor electronics. A key obstacle to integrating magnonic functionality with electronic hardware, however, remains the lack of scalable nanoscale electrical interfaces for magnon readout.

Magnon dynamics can be resolved directly using synchrotron-based techniques such as scanning transmission X-ray microscopy\cite{wintzMagneticVortexCores2016} or detected through the dynamic stray magnetic fields generated by the precessing magnetisation. Electrical approaches include inductive antennas\cite{vlaminckCurrentInducedSpinWaveDoppler2008} and magnetoresistive sensors based on the giant magnetoresistance effect\cite{rossiMagnetoresistiveDetectionSpin2025}, while nanoscale quantum spin sensing provides high spatial resolution but relies on optical readout\cite{vandersarNanometrescaleProbingSpin2015}. These techniques have established powerful methods for studying magnon dynamics, but combining nanoscale electrical detection with a device technology that is compatible with dense, wafer-scale semiconductor fabrication remains challenging.

Here, we demonstrate electrical magnon spectroscopy using a 70-\SI{}{\nano\meter}-diameter magnetic tunnel junction (MTJ) derived from magnetoresistive random-access memory (MRAM) technology and fabricated as part of a dense array on a commercial 300-\SI{}{\milli\meter} wafer using the GlobalFoundries 22FDX$^\textsuperscript{\textregistered}$ process.
The MTJ consists of two ferromagnetic layers separated by a nonmagnetic tunnel barrier, with its electrical resistance determined by their relative magnetisation orientation through tunnelling magnetoresistance (TMR)\cite{yuasaGiantRoomtemperatureMagnetoresistance2004,parkinGiantTunnellingMagnetoresistance2004}. In conventional MRAM, stable parallel and antiparallel configurations encode binary information\cite{gallagherDevelopmentMagneticTunnel2006,kentNewSpinMagnetic2015,worledgeSpintransferTorqueMagnetoresistive2024}. Here, we instead engineer the magnetic anisotropy of the free layer such that the MTJ acts as a sensitive nanoscale detector of dynamic magnetic fields. Coupling the sensor to a magnetic vortex allows the dipolar fields generated by magnons to be transduced into microwave-frequency electrical signals.
By exploiting three-magnon splitting, we spectrally separate these signals from the direct microwave excitation and electrically resolve the complex nonlinear magnon dynamics previously accessible primarily through Brillouin light-scattering spectroscopy\cite{schultheissExcitationWhisperingGallery2019,korberNonlocalStimulationThreeMagnon2020,korberModificationThreemagnonSplitting2023,merboucheDegenerateNondegenerateParametric2024}. 
Importantly, the electrical readout reveals spectral features that remain unresolved in the optical measurements, including nonlinear magnon modes with linewidths down to 188 kHz. Together, these results establish foundry-fabricated MTJs as nanoscale electrical interfaces for high-resolution magnon spectroscopy and provide a route towards integrating magnonic functionality with semiconductor electronics.

\begin{figure*}
    \includegraphics{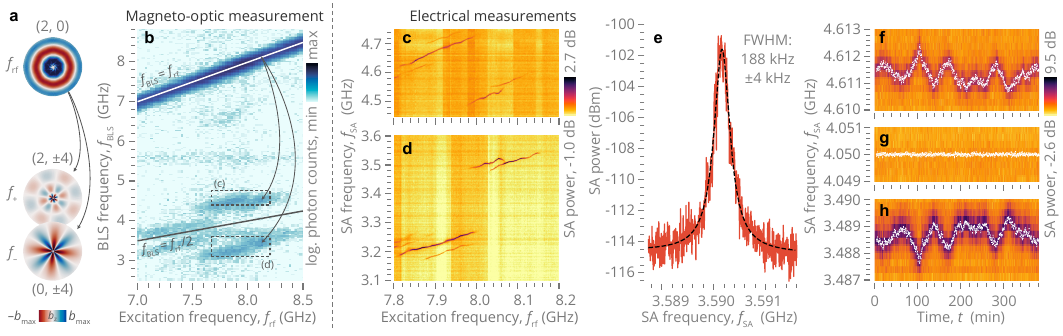}
     \caption{\textbf{Electrical spectroscopy of nonlinear magnon dynamics.} \textbf{a}, Exemplary three-magnon splitting process. \textbf{b},  Magneto-optical Brillouin light scattering (BLS) spectra recorded as a function of the excitation frequency, with the detected intensity colour-coded. \textbf{c,d}, Electrically detected magnon spectra recorded with the MTJ in the frequency windows indicated by dashed rectangles in \textbf{b}. Restricting the frequency range to the nonlinear split modes separates the magnon signals from the drive frequency, where direct microwave cross-talk can contribute. \textbf{e}, High-resolution electrical spectrum revealing a linewidth of \SI{188}{\kilo\hertz} at a magnon frequency of \SI{3.59}{\giga\hertz}. \textbf{f},\textbf{h}, Frequencies of the two magnon modes generated by three-magnon splitting, recorded electrically over time under continuous excitation at $f_\text{rf}=\SI{8.1}{\giga\hertz}$. \textbf{g}, Mean frequency $(f_+ + f_-)/2$ of the two split modes, remaining locked to half the excitation frequency $f_\text{rf}/2$. }
    \label{fig:exp}
\end{figure*}

\section*{Foundry-fabricated MTJ sensor for magnon readout}

We demonstrate nanoscale magnon readout in a model system comprising a ferromagnetic Ni$_{81}$Fe$_{19}$ disk with a diameter of \SI{5}{\micro\meter} and a thickness of \SI{50}{\nano\meter}. At equilibrium, the ferromagnet forms a magnetic vortex state\cite{cowburnSingleDomainCircularNanomagnets1999, shinjo_magnetic_2000}, which is stable at zero magnetic field and room temperature (Fig.~\ref{fig:platform}a). The disk acts as a cavity for magnon resonances with cylindrical symmetry, which are geometrically quantised and characterised by their radial and azimuthal mode numbers ($n$, $m$)\cite{buessExcitationsNegativeDispersion2005,ivanovHighFrequencyModes2005,vogtOpticalDetectionVortex2011}. The overlay in Fig.~\ref{fig:platform}b illustrates the amplitude profile of mode ($0, 1$) across half the disk. 

When magnons are excited, the magnetic moments precess out of the film plane, creating alternating magnetic surface charges, as schematised in Fig.~\ref{fig:platform}b. These charges oscillate in time at the magnon frequency (1 to \SI{20}{\giga\hertz}) and in space on the scale of the magnon wavelength (from a few micrometres down to hundreds of nanometres), thereby generating dynamic dipolar fields above and below the disk. 
Figure~\ref{fig:platform}c shows the $z$-component of the dipolar field generated by mode ($0, 4$) at different distances $|z|$ from the disk surface. The field amplitude ($b_\text{max}$) decays rapidly with distance, and its spatial features become increasingly blurred on length scales comparable to the magnon wavelength. 
These dynamic dipolar fields provide the transduction mechanism: at the position of the MTJ, they drive the magnetisation of its free layer and thereby convert the magnon dynamics into an electrical signal.

To realise this transduction experimentally, we fabricate a hybrid structure in which the magnetic disk is positioned above a customised MRAM array, as sketched in Fig.~\ref{fig:platform}d. The starting point is a full 300-\SI{}{\milli\meter} wafer processed on the 22FDX$^\textsuperscript{\textregistered}$ technology at GlobalFoundries, Dresden. Each array contains about 20,000 circular MTJs, approximately \SI{70}{\nano\meter} in diameter and separated by an edge-to-edge distance of \SI{150}{\nano\meter}. The tunnelling magnetoresistance stack is based on a CoFeB/MgO/CoFeB multilayer, with the free layer located approximately \SI{200}{\nano\meter} below the wafer surface. 

Processing is subsequently continued on individual $\SI{10}{\milli\meter} \times \SI{10}{\milli\meter}$ chips. Following deposition of a 50-\SI{}{\nano\meter}-thick SiO$_2$ insulation layer, the magnetic disk is patterned above the MRAM array. The free layer of the MTJ is therefore located approximately \SI{250}{\nano\meter} below the magnetic disk, corresponding to the experimentally relevant distance considered in Fig.~\ref{fig:platform}c. A gold $\Omega$-shaped microwave antenna surrounding the disk generates oscillatory magnetic fields to excite the magnons. 
Figure~\ref{fig:platform}e shows a focused-ion-beam cross-section of the resulting hybrid structure imaged by scanning electron microscopy (SEM) at an angle of 52$^\circ$. The lower part of the image reveals a row of buried MRAM cells connected by bit and source lines, while the upper surface shows the magnetic disk surrounded by the $\Omega$-shaped antenna.

The test chips provide electrical access to a selected MTJ through one bit line and one source line connected to larger contact pads, as highlighted in Fig.~\ref{fig:platform}e. This architecture separates excitation and detection: the microwave antenna excites collective magnon modes in the magnetic disk, while the selected MTJ locally detects the dynamic dipolar fields generated by these modes. The MTJ is therefore used here as an individual nanoscale sensor within a densely fabricated MRAM array. A transmission electron microscopy (TEM) image of a customised MTJ is shown in Fig.~\ref{fig:platform}f. The MgO tunnel barrier is clearly resolved, while the ferromagnetic layers are embedded between the seed and capping layers connecting the MTJ to the bit line along the $x$-direction and the source line along the $y$-direction. 

To optimise the MTJs for magnetic-field sensing, we engineer the MgO–free-layer interface and increase the free-layer thickness compared with a conventional MRAM stack. This reduces the effective perpendicular magnetic anisotropy such that the free-layer magnetisation lies in the sample plane at zero field while remaining highly susceptible to small out-of-plane magnetic fields. The magnetoresistance measured as a function of the out-of-plane field $B_\text{z}$ in Fig.~\ref{fig:platform}g corroborates this behaviour and demonstrates the easy transition between the parallel and antiparallel resistance states at negative and positive fields, respectively.
Importantly, the magnon frequencies investigated here are spectrally separated from the intrinsic normal modes of the free layer, such that the MTJ provides a non-resonant response to the magnon-induced dipolar fields rather than acting as an independently resonant microwave element.

Figure~\ref{fig:platform}h illustrates the resulting free-layer dynamics at two instances separated by half a precession cycle. The oscillatory response generates parallel and antiparallel magnetisation components $\pm\Delta m_\text{z}$ relative to the reference layer. The tunnelling magnetoresistance therefore oscillates at the magnon frequency and can be measured electrically with a spectrum analyser connected to the MTJ.

\section*{Spectral separation of magnon signal and microwave input}

A major challenge in electrical magnon detection is separating magnon-generated signals from the direct microwave input. Magnons are typically excited by oscillating magnetic fields in the mT range generated by microwave antennas. These fields can also couple directly to the free-layer magnetisation of the MTJ, while electromagnetic cross-talk between the antenna and the sensor's bit and source lines can generate electrical signals at the excitation frequency. A response detected at the drive frequency alone therefore cannot be unambiguously attributed to magnons.

We overcome this ambiguity by exploiting nonlinear magnon scattering in the vortex-state disk as a frequency-conversion mechanism. Specifically, we use three-magnon splitting\cite{schultheissExcitationWhisperingGallery2019,verbaTheoryThreemagnonInteraction2021}, in which a strongly excited magnon mode decays into two secondary modes at lower frequencies. An exemplary splitting process is illustrated in Fig.~\ref{fig:exp}a. Energy conservation requires $f_\text{rf} = f_+ + f_-$, while conservation of angular momentum imposes the corresponding selection rule $m_\text{rf} = m_- + m_+$. The process is nonlinear and occurs only above a well-defined excitation threshold.

Crucially, the secondary magnons appear at frequencies distinct from the applied microwave drive. Detecting these modes electrically therefore provides a spectrally separated magnon signal that can be distinguished from direct microwave excitation and feed-through. This nonlinear frequency conversion enables the MTJ to probe the intrinsic magnon dynamics of the vortex while suppressing ambiguity from the excitation circuitry.

\begin{figure*}
    \includegraphics{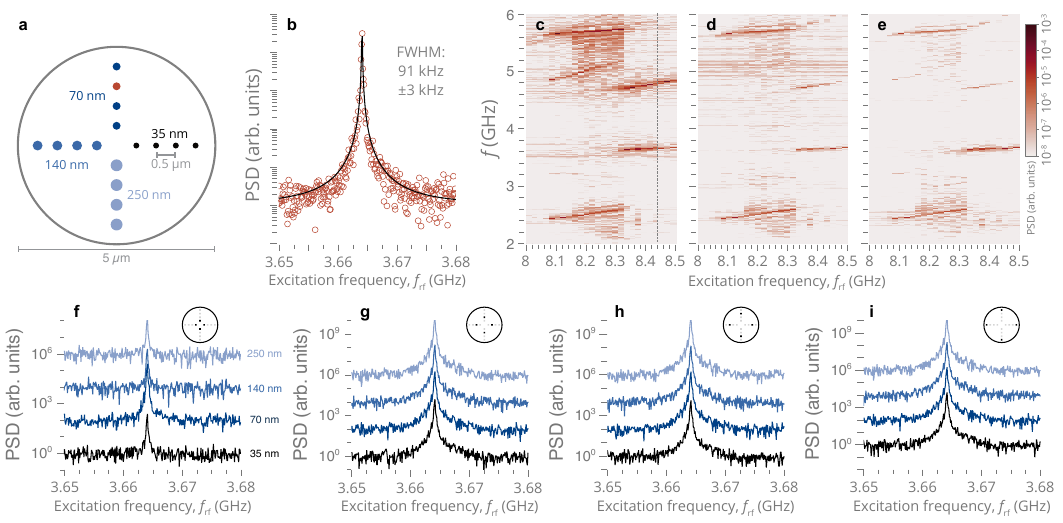}
     \caption{\textbf{Simulated power spectra of scattered modes.} \textbf{a}, Disk geometry and simulated sensor positions, located \SI{250}{\nano\meter} above the ferromagnetic film surface, at which the dynamic dipolar stray fields are calculated. The orange point represents a circular region with a diameter of \SI{70}{\nano\meter} at a radial distance of \SI{1.5}{\micro\meter} from the disk centre. \textbf{b}, Simulated power spectral density (PSD) of the $x$-component dynamic dipolar fields at the orange point in \textbf{a}, around the scattered mode $f_{-}$, for an excitation frequency of \SI{8.44}{\giga\hertz} and an amplitude of \SI{2}{\milli\tesla}. Circles represent simulation data and the solid black curve is a Lorentzian fit. \textbf{c-e}, Colour map of the simulated PSD of the \textbf{c}, $x$-, \textbf{d}, $y$-, and \textbf{e}, $z$-components of the dynamic dipolar field at the orange point in \textbf{a} as a function of the excitation frequency $f_\mathrm{rf}$, for an excitation amplitude of \SI{2}{\milli\tesla}. The dashed line in \textbf{c} indicates the value of $f_\mathrm{rf}$ in \textbf{b}. \textbf{f}-\textbf{i}, Simulated PSD of the $x$-component of the dynamic dipolar field for the mode $f_{-}$ shown in \textbf{b}, but for different sensor sizes and radial positions as indicated in \textbf{a}: \textbf{f}, \SI{0.5}{\micro\meter}, \textbf{g}, \SI{1.0}{\micro\meter} \textbf{h}, \SI{1.5}{\micro\meter}, and \textbf{i}, \SI{2.0}{\micro\meter}. A vertical offset is applied to the curves to facilitate comparisons.}
    \label{fig:sim}
\end{figure*}

\section*{Electrical characterisation of magnon spectra}

We first characterise the directly excited and scattered magnon spectra using using magneto-optical Brillouin light scattering (BLS) microscopy\cite{sebastian_micro-focused_2015} (see Methods for details). Figure~\ref{fig:exp}b shows the measured spectra as a function of the excitation frequency $f_\text{rf}$ applied to the $\Omega$-shaped microwave antenna. The diagonal response at $f_\text{BLS}=f_\text{rf}$ corresponds to the directly excited mode, while the off-diagonal branches appear in pairs around $f_\text{rf}/2$ and result from three-magnon splitting.

We then perform electrical spectroscopy of the same nonlinear magnon modes using the MTJ. For each excitation frequency, the MTJ is contacted with a high-frequency ground-signal-ground (GSG) probe and its power spectrum is recorded with a spectrum analyser (see Methods for details). The oscillatory magnetisation of the free layer is detected through the resulting modulation of the tunnelling magnetoresistance. Figures~\ref{fig:exp}c,d show the electrical spectra within the frequency windows of the split modes indicated by the dashed rectangles in Fig.~\ref{fig:exp}b. Restricting the measurement to these frequency ranges separates the magnon-generated signals from the drive frequency, where direct microwave cross-talk can contribute.

The electrically detected branches reproduce the spectral evolution observed by BLS, but with markedly higher frequency resolution. Because the individual electrical resonances are much narrower than the optically resolved features, they appear only faintly in the colour maps. For visibility, the spectra plotted in Fig.~\ref{fig:exp}c,d are therefore convoluted with a Gaussian function with a full width at half-maximum (FWHM) of \SI{1}{\mega\hertz}. A representative unconvolved spectrum is shown in Fig.~\ref{fig:exp}e. A Lorentzian fit yields a linewidth of \SI{188}{\kilo\hertz} at a centre frequency of \SI{3.59}{\giga\hertz}. The signal occurs at the frequency of a secondary magnon generated by three-magnon splitting and is therefore spectrally distinct from the applied microwave drive.

The high spectral resolution of the electrical readout reveals a richer nonlinear magnon spectrum than is resolved by BLS. Features that appear as a single broad resonance in Fig.~\ref{fig:exp}b separate into multiple distinct modes in Fig.~\ref{fig:exp}c,d, with frequency separations several times larger than their individual linewidths. The MTJ readout also allows the mode frequencies to be tracked over time with high precision. Figures~\ref{fig:exp}f,h show the frequencies of the two split modes under continuous excitation at $f_\text{rf}=\SI{8.1}{\giga\hertz}$, with each power spectrum acquired over \SI{23.1}{\second}. Both frequencies fluctuate by several megahertz, whereas their mean $(f_+ + f_-)/2$, shown in Fig.~\ref{fig:exp}g, remains locked to $f_\text{rf}/2$, as required by energy conservation. Together with the stable spatial mode profiles observed over several hours in BLS measurements\cite{schultheissExcitationWhisperingGallery2019}, this indicates that the frequency variations arise from slow shifts of the magnon spectrum, which we attribute to temperature drift, rather than from mode hopping.

\section*{Micromagnetic simulations}

To further examine the origin of the electrically detected signals, we perform finite-temperature micromagnetic simulations of three-magnon scattering under experimental conditions (see Methods). The simulations allow us to determine the dynamic dipolar fields generated by the nonlinear split modes at the position of the MTJ and to compare their spectral characteristics with the electrical measurements.

The narrow linewidths of the scattered modes, at least two orders of magnitude below those expected for freely damped oscillations, are consistent with self-sustained oscillations subject to thermal noise, as observed in spin-torque nano-oscillators\cite{kimGenerationLinewidth2008,kimSpinTorque2012}. In this picture, each scattered mode $f_-$ and $f_+$ behaves as a limit-cycle oscillator sustained by energy transfer from the directly excited mode through three-magnon scattering.

We first evaluate the dynamic stray field at different radial distances from the centre of the ferromagnetic disk and average it over circular areas corresponding to different MTJ diameters, as illustrated in Fig.~\ref{fig:sim}a. The experimentally relevant configuration is represented by the orange marker, corresponding to a \SI{70}{\nano\meter}-diameter sensor positioned at a radial distance of \SI{1.5}{\micro\meter} from the disk centre and approximately \SI{250}{\nano\meter} from the magnetic film. Figure~\ref{fig:sim}b shows the simulated power spectrum of a scattered mode at this position for an excitation frequency of \SI{8.44}{\giga\hertz}. The spectrum is calculated solely from the dynamic dipolar field and exhibits a linewidth of \SI{91}{\kilo\hertz}, comparable to the experimentally observed sub-MHz linewidths.

Figures~\ref{fig:sim}c-e show the simulated power spectra of the $x$-, $y$-, and $z$-components of the dynamic dipolar field as the excitation frequency is varied. All three components exhibit similar branches of the nonlinear split modes to those observed experimentally in Fig.~\ref{fig:exp}. This indicates that the characteristic spectral response is a property of the magnon dynamics in the vortex rather than being restricted to a particular orientation of the MTJ free-layer magnetisation.

Finally, Figs.~\ref{fig:sim}f-i compare the simulated spectra for the different sensor sizes and radial positions shown in Fig.~\ref{fig:sim}a. Apart from the reduced signal strength close to the vortex centre, the characteristic spectral response is robust against variations in sensor size and position. The simulations therefore support the interpretation that the MTJ acts as a local transducer of the dynamic dipolar fields generated by the nonlinear magnon modes, while the observed spectral structure is governed primarily by the magnetisation dynamics of the vortex.

\section*{Conclusion}

We have demonstrated electrical magnon spectroscopy using a nanoscale MTJ derived from MRAM technology and fabricated on a commercial 300-\SI{}{\milli\meter} semiconductor wafer. Dynamic dipolar fields generated by magnons in a magnetic vortex are transduced into electrical signals through the tunnelling magnetoresistance of a \SI{70}{\nano\meter}-diameter sensor. By exploiting three-magnon splitting, the magnon-generated signals are spectrally separated from the microwave drive, enabling unambiguous electrical detection of nonlinear magnon dynamics. The MTJ readout resolves spectral features that remain unresolved in the parallel BLS measurements, including magnon linewidths down to \SI{188}{\kilo\hertz} at \SI{3.59}{\giga\hertz}.

The use of an MRAM-derived sensing element provides a direct link between magnonic functionality and an established, wafer-scale spintronic technology. The present device operates at room temperature and without an external static magnetic field, while the dense MTJ array provides a natural basis for future architectures with multiple electrically addressable sensors. Such layouts could enable parallel or spatially resolved electrical access to magnonic circuits while retaining the local dipolar-field transduction mechanism demonstrated here. More broadly, the sub-MHz linewidths revealed by the electrical spectroscopy may also be relevant for hybrid magnonic systems, where narrow magnon resonances are beneficial for coherent coupling to other excitations. Our results therefore establish foundry-fabricated MTJs as nanoscale electrical interfaces for magnon spectroscopy and provide a path towards integrating magnonic signal processing with semiconductor electronics.


\section*{Author declarations}

\subsection*{Conflict of Interest}
The authors have no conflicts of interest to disclose.

\subsection*{Author contributions}

H.S., K.S., and C.H. conceived the experiments. K.S., Z.X., R.N., A.T., and J.M. fabricated the samples. C.H. carried out the experiments. T.D. assisted with the experimental setup. J.-V.K. and A.K. performed micromagnetic simulations. R.H. and A.W. prepared and recorded the SEM and TEM images. C.H., J.-V.K, K.S., and H.S. visualised the results. J.-V.K., T.D., A.K., K.S., J.F., A.T., J.M., and H.S. acquired funding. All authors analysed the data and discussed the results. H.S., K.S., and J.-V.K. wrote the original draft of the paper. All authors reviewed and edited the paper.


\section*{Acknowledgments}

Support by the Nanofabrication Facilities Rossendorf (NanoFaRo) and the Structural Characterization Facilities Rossendorf at the IBC are gratefully acknowledged. The project received financial support from the EU Research and Innovation Programme Horizon Europe under grant agreement no. 101070290 (NIMFEIA). This work was supported by the European Union and the Free State of Saxony within the projects MagKI and Magnon4KI.

\section*{Data availability}
The data that support the findings of this study are openly available in RODARE\cite{heins_christopher_2025_3988}.


\bibliographystyle{naturemag}
\bibliography{references.bib}

\section*{Methods} 

\subsection*{Sample fabrication}

The samples were fabricated starting from a full 300-\SI{}{\milli\meter} wafer processed using the 22FDX$^\textsuperscript{\textregistered}$ technology at GlobalFoundries, Dresden. Each die contained an MRAM array comprising approximately 20,000 circular MTJs, each about \SI{70}{\nano\meter} in diameter and separated by an edge-to-edge distance of \SI{150}{\nano\meter}. The tunnelling magnetoresistance stack is based on a CoFeB/MgO/CoFeB multilayer. To minimise the separation between the MTJ and the subsequently fabricated magnon cavity, the wafer surface above the MRAM array was polished at GlobalFoundries, reducing the distance between the wafer surface and the MTJ free layer to approximately \SI{200}{\nano\meter}.

Further processing was performed on individual $\SI{10}{\milli\meter}\times\SI{10}{\milli\meter}$ chips at the Helmholtz-Zentrum Dresden-Rossendorf using three lithography steps. First, a \SI{50}{\nano\meter}-thick SiO$_2$ insulation layer was patterned using a double-layer resist of LOL-2000 and S1805, direct-write laser lithography, electron-beam evaporation, and lift-off. The resulting separation between the magnetic disk and the MTJ free layer is therefore approximately \SI{250}{\nano\meter}.

Second, the \SI{5}{\micro\meter}-diameter magnetic disk was patterned above the MRAM array using the same double-layer resist and direct-write laser lithography, followed by electron-beam evaporation of Cr(\SI{3}{\nano\meter})/Ni${80}$Fe${20}$(\SI{50}{\nano\meter})/Cr(\SI{3}{\nano\meter}) and lift-off.

\begin{figure}
    \includegraphics{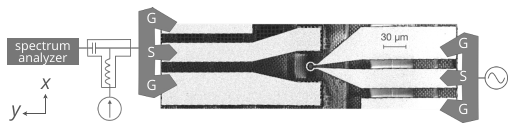}
     \caption{\textbf{Electrical contacting.} Scanning electron microscopy image showing the top view of the sample including coplanar wave-guides for contacting the MRAM cell and microwave antenna, respectively.}
    \label{fig:sample}
\end{figure}

In a third lithography step, the $\Omega$-shaped microwave antenna, coplanar waveguides, and electrical contact structures were fabricated by electron-beam lithography. A double-layer resist of ethyl lactate (EL11) and polymethyl methacrylate (950PMMA A4) was patterned, followed by electron-beam evaporation of Cr(\SI{5}{\nano\meter})/Au(\SI{150}{\nano\meter}) and lift-off. The inner and outer diameters of the $\Omega$-shaped antenna are \SI{8}{\micro\meter} and \SI{11}{\micro\meter}, respectively. The antenna generates the oscillatory magnetic field used to excite magnons in the disk, while the coplanar waveguides connect the antenna and the electrically accessible MTJ to microwave-frequency ground-signal-ground (GSG) contact pads, as shown in the scanning electron microscopy image in Fig.~\ref{fig:sample}.

Because the magnetic disk and the Au antenna are defined in separate lithography steps, a small lateral offset between the two structures is visible in Fig.~1e and results from the finite overlay accuracy of the fabrication process. This offset modifies the spatial profile of the excitation field and can therefore influence the relative excitation efficiencies of individual magnon modes, but does not alter the underlying nonlinear scattering process. 

The coplanar waveguides were designed to position the GSG probes sufficiently far from the magnetic disk to permit simultaneous electrical and magneto-optical measurements. This geometry allows the disk to be accessed with a high-numerical-aperture microscope objective while the MTJ and microwave antenna remain electrically contacted.

\subsection*{Electrical readout of MTJ }

For the electrical measurements, the MTJ is contacted with a GSG probe and connected to the combined port of a high-frequency bias tee. A dc bias current of \SI{150}{\micro\ampere} is supplied through the dc port using a Keithley 2614B. No microwave current is intentionally applied through the MTJ. Instead, the time-dependent tunnelling magnetoresistance generated by the dynamic response of the free layer produces a microwave voltage across the dc-biased junction. The rf port of the bias tee is connected through a low-noise amplifier (Mini-Circuits ZX60-83LN-S+) to a spectrum analyser (Keysight N9010A). The measured microwave signal reflects the response of the MTJ free layer to the local dynamic magnetic field. 
 
For background subtraction, a reference spectrum is recorded with the same dc bias current but without microwave excitation applied to the antenna. The spectra in Figs.~\ref{fig:exp}c,d,f-h are recorded with a resolution bandwidth (RBW) of \SI{150}{\kilo\hertz}. For the high-resolution spectrum in Fig.~\ref{fig:exp}e, the RBW is reduced to \SI{2}{\kilo\hertz}. All spectra are acquired using signal averaging.

\subsection*{Brillouin light scattering}

All experimental measurements were performed at room temperature. The magnon spectra were detected by means of Brillouin light scattering microscopy~\cite{sebastian_micro-focused_2015}. Therefore, a monochromatic, continuous-wave 532-\SI{}{\nano\meter} laser was focused onto the sample surface using a microscope lens with a high numerical aperture, yielding a spatial resolution of about \SI{300}{\nano\meter}. The backscattered light was then directed into a Tandem Fabry-P\'{e}rot interferometer~\cite{mock_construction_1987} in order to measure the frequency shift caused by the inelastic scattering of photons and magnons. The detected intensity of the frequency-shifted signal is directly proportional to the magnon intensity at the respective focusing position. 

During all experiments, the investigated  microstructure was imaged using a red LED and a CCD camera. Displacements and drifts of the sample were tracked by an image recognition algorithm and compensated by the sample positioning system.

To account for the different spatial distributions of the magnon modes, the signal was integrated over 3 radial and 4 azimuthal positions across half the 5-µm-wide disk.

\subsection*{Micromagnetic simulations}

The micromagnetic simulations were performed with the open-source \texttt{MuMax3} code~\cite{vansteenkiste_design_2014, leliaert_adaptively_2017}, which uses the finite-difference method to solve the Landau-Lifshitz equation with Gilbert damping,
\begin{equation}
\frac{d\mathbf{m}}{dt} = -|\gamma_0| \mathbf{m} \times \mathbf{H}_\mathrm{eff} + \alpha \mathbf{m} \times \frac{d\mathbf{m}}{dt},
\end{equation}
where $\| \mathbf{m}(\mathbf{r},t) \| =1$ is a unit vector representing the orientation of the magnetisation field, $\gamma_0 = \mu_0 \gamma$ is the gyromagnetic constant, and $\alpha$ is the Gilbert damping constant. With $M_s$ denoting the saturation magnetisation, the effective field, $\mathbf{H}_\mathrm{eff} = -(1/\mu_0 M_s) \delta U/\delta \mathbf{m}$, represents a variational derivative of the total magnetic energy $U$ with respect to the magnetisation, with $U$ containing contributions from the Zeeman, nearest-neighbour Heisenberg exchange, and dipolar interactions. We account for finite temperatures by supplementing the effective field with a random thermal field, $\mathbf{H}_\mathrm{eff} \rightarrow \mathbf{H}_\mathrm{eff} + \mathbf{h}_\mathrm{th}$. This random field represents a Gaussian white noise with the spectral properties
\begin{align}
\langle h_{\mathrm{th},i}(\mathbf{r},t) \rangle &= 0, \\
\langle h_{\mathrm{th},i}(\mathbf{r},t) h_{\mathrm{th},j}(\mathbf{r}', t') \rangle &= q \, \delta_{ij}\delta(\mathbf{r}-\mathbf{r}')\delta(t-t'),
\end{align}
where $q = 2 \alpha k_B T/(\mu_0 \gamma_0 M_s)$ and $i,j=x,y,z$ represent the different Cartesian components of the field. We assumed micromagnetic parameters consistent with permalloy, with $M_s = 765$~kA/m, an exchange constant of $A = 12$~pJ/m, a Gilbert damping constant of $\alpha = 0.008$, $\gamma = 1.86 \times 10^{11}$~rad/s, and a temperature of $T=300$~K. The simulated 5-$\mu$m diameter, 50-nm thick disk is represented within a rectangular box with dimensions of 5~$\mu$m $\times$ 5~$\mu$m $\times$ 50~nm, which is discretised using $1024 \times 1024 \times 1$ finite difference cells. 

We model the power spectrum of the MRAM sensors by assuming that the magnetoresistance signal arises purely from the dynamic response of the free layer of the MTJ to the dynamic dipolar fields $\mathbf{h}_d$ generated by the magnetisation processes within the ferromagnetic film. Provided that the relevant magnons at play within the Permalloy disk do not coincide with the frequencies of the resonant modes of the MTJ free layer, we can assume that the free layer magnetisation can be expressed as $m_{\mathrm{FL},i}(t) = \chi_{ij} h_{d,j}(t)$, where $\chi$ is a magnetic susceptibility with $\mathrm{Re}(\chi_{ij}) \neq 0$ and $\mathrm{Im}(\chi_{ij})=0$. As such, we assume that the power spectrum of the free layer magnetisation fluctuations 
\begin{equation}
S_\mathrm{FL}(\omega) = \int_{-\infty}^{\infty} \langle m_{\mathrm{FL}}(t) m_{\mathrm{FL}}(t-t') \rangle e^{-i \omega t'} dt'  
\end{equation}
can be expressed in terms of the power spectrum of the dynamic dipolar fields seen by the MTJ free layer,
\begin{equation}
S_\mathrm{FL}(\omega) \propto \int_{-\infty}^{\infty} \langle h_{d}(t) h_{d}(t-t') \rangle e^{-i \omega t'} dt'.
\end{equation}

We compute the dynamic dipolar fields at a distance of 250~nm above the ferromagnetic film surface, averaged over the volumes of the different cylinders in Fig.~\ref{fig:sim}(a), which represent different MTJ sensor positions and sizes. For each of the power spectra displayed in Fig.~\ref{fig:sim}, we perform time-integration of the stochastic Landau-Lifshitz equation over an interval of $t_0$, recording $h_d$ every 20~ps. We then estimate the power spectrum for each Cartesian component of the dynamic dipolar field as
\begin{equation}
S_i(\omega) \simeq \left| \int_{0}^{t_0} h_{d,i}(t) e^{-i \omega t} dt \right|^2.
\end{equation}
For Figs.~\ref{fig:sim}(b,f,g,h,i), $S_x(\omega)$ is averaged over three simulations with $t_0 =$ 10~$\mu$s in which each simulation used a different realisation of the random field $h_\mathrm{th}$, while for Figs.~\ref{fig:sim}(c,d,e), $S_i(\omega)$ is computed from a single simulation over $t_0 =$ 200~ns for each value of the excitation frequency. We model the experimental excitation field by including a spatially-uniform sinusoidal field $\mathbf{h}_\mathrm{rf} = h \sin (2\pi f_\mathrm{rf}t) \hat{\mathbf{z}}$ along the direction perpendicular to the film plane.

The spatial mode profiles of the three magnon splitting process in Fig.~\ref{fig:exp}(a) are found by applying a sinusoidal out-of-plane (OOP) magnetic field at $f_\mathrm{rf}=\SI{8.27}{\giga\hertz}$ with an amplitude of \SI{0.25}{\milli\tesla} for a total duration of \SI{100}{\nano\second}.
The time-dependent magnetisation was sampled at a rate
of \SI{10}{\pico\second} and subsequently Fourier-transformed in the time domain at each cell. The peak positions of the satellites $f_-$ and $f_+$ were determined and the respective mode amplitudes were calculated via inverse Fourier transform at these frequencies. 

For the split mode $f_-$, the dipolar fields were computed and are shown in Fig.~\ref{fig:platform}(c). The mode was excited with an OOP field of matching symmetry,  
$\mathbf{h}_\mathrm{rf} = h \, \sin(2\pi f_\mathrm{rf} t)\, \cos(m\phi)\, \hat{\mathbf{z}},$
where $m=4$, $h=\SI{0.5}{\milli\tesla}$, and $\phi = \operatorname{atan2}(y,x)$. Calculations were performed in three planes beneath the disk with a discretisation of $512 \times 512 \times 1$ cells each.

\end{document}